\documentclass[aps,pra,twocolumn,groupedaddress]{revtex4-1}
\usepackage{amsmath}
\usepackage{graphicx}
\usepackage{dcolumn}
\usepackage[ruled]{algorithm2e}
\usepackage{bm}
\usepackage[T1]{fontenc}
\usepackage{mathptmx}
\usepackage[utf8]{inputenc}
\usepackage{mathtools}
\usepackage{mathrsfs}
\usepackage{amssymb}
\usepackage{amsfonts}
\usepackage{graphicx}
\usepackage{braket}
\usepackage{float} 
\usepackage{color,xcolor}
\usepackage{booktabs}
\usepackage{subcaption}
\usepackage{diagbox}
\usepackage[marginal]{footmisc}
\usepackage{textcomp}
\usepackage[colorlinks,linkcolor=blue]{hyperref}
\usepackage{lipsum}
\begin{document}

\title{Generic security analysis framework for quantum secure direct communication}
\author{Zhangdong Ye$^{1}$}
\author{Dong Pan$^1$}
\author{Zhen Sun$^3$}
\author{Chunguang Du$^1$}
\author{Liuguo Yin$^{2,3,4,5}$}
\email{yinlg@tsinghua.edu.cn}
\author{Guilu Long$^{1,2,4,5}$}
\email{gllong@tsinghua.edu.cn}

\affiliation{$^{1}$ State Key Laboratory of Low-Dimensional Quantum Physics and Department of Physics, Tsinghua University, Beijing 100084, China}
\affiliation{$^{2}$ Frontier Science Center for Quantum Information, Beijing 100084, China}
\affiliation{$^{3}$ School of Information and Technology, Tsinghua University, Beijing 100084, China}
\affiliation{$^{4}$ Beijing National Research Center for Information Science and Technology, Beijing 100084, China}
\affiliation{$^{5}$ Beijing Academy of Quantum Information Sciences, Beijing 100193, China}

\date{\today}
\begin{abstract}
Quantum secure direct communication provides a direct means of conveying secret information via quantum states among legitimate users. The past two decades have witnessed its great strides both theoretically and experimentally. However, the security analysis of it still stays in its infant. Some practical problems in this field to be solved urgently, such as detector efficiency mismatch, side-channel effect and source imperfection, are propelling the birth of a more impeccable solution. In this paper, we establish a new framework of the security analysis driven by numerics where all the practical problems may be taken into account naturally. We apply this framework to several variations of the DL04 protocol considering real-world experimental conditions. Also, we propose two optimizing methods to process the numerical part of the framework so as to meet different requirements in practice. With these properties considered, we predict the robust framework would open up a broad avenue of the development in the field.
\end{abstract}

\maketitle

%\onecolumngrid

\section{Introduction}
\noindent Quantum secure direct communication (QSDC) was proposed by Long and Liu in 2000~\cite{long2002theoretically,long2007quantum}, which is a way of achieving secure communication by transmitting secret information directly over the quantum channel. Guaranteed by quantum-mechanical properties of the information carriers, say entangled photons~\cite{long2002theoretically,deng2003two,wang2005quantum} or single photons~\cite{deng2004secure}, two legitimate distant parties can detect eavesdropping on-site during the communication via random sampling of the quantum states. The past two decades have witnessed the blossom of QSDC both theoretically and experimentally. In addition to point-to-point protocols~\cite{long2002theoretically,deng2003two,wang2005quantum,deng2004secure}, multiuser communication schemes have also made great strides~\cite{deng2006quantum,fu2007economical}. Recently, the theoretical protocols of measurement-device-independent QSDC that eliminate the loopholes of the measurement devices have been proposed~\cite{zhou2020measurement,niu2018measurement,gao2019long,Zou_2020,Wu2020}, while device-independent QSDC protocols that relax the security assumptions on the quantum devices are brewing up for example in Ref.~\cite{zhou2019device}. Meanwhile, more interesting schemes contributed to the aim of QSDC have been established, such as quantum illumination~\cite{shapiro2014secure}, quantum data locking~\cite{lum2016quantum} and quantum low probability of intercept~\cite{shapiro2019quantum}. In the aspect of experiments, the first proof-of-principle implementation using a frequency coding strategy~\cite{hu2016experimental} demonstrates the feasibility of QSDC over a noisy quantum channel, which is afterwards followed by a demonstration experiment of entanglement-based QSDC protocol materialized by the quantum-memory-assisted (QMA) system~\cite{zhang2017quantum}. In particular, the QMA system makes it promising to conduct super-long-distance communication~\cite{zhu2017experimental} and to construct QSDC networks. 
The free-space communication scheme has been studied as well, shown in the literature~\cite{pan2020experimental}. Moreover, some typical applications of optical quantum information have been presented~\cite{5175331,liu2019single,PhysRevResearch.1.033063,PMID:31912033}, which are promisingly potential to facilitate the implementation of QSDC.

Despite the great progress achieved, the security analysis of QSDC had been staying at the qualitative stage for a time before Qi, \textit{et al} came up with the first quantitative analysis framework~\cite{qi2019implementation} illuminated by the two-way QKD analysis strategy in Refs.~\cite{Hua2011,Ivan2015}. On the top of Qi's framework, the work in Ref.~\cite{wu2019security} gives a further exposition on the asymptotic secrecy capacity of QSDC under the collective attacks. However, some idealized assumptions have to be made in this framework to accommodate the strategy used in Ref.~\cite{Ivan2015}. For example, bits "0" and "1" come up randomly in the encoded message and furthermore, the information source could be perfectly compressed. On the other hand, the calculation to find the eigenvalues of the Gram matrix involved is pretty mathematically technical especially when the composite system of the legitimate users and the adversary becomes complicated in the cases where practical conditions are considered or higher dimensional protocols are carried out.

In this work, we establish a new framework of the security analysis to completely address the above-stated problems getting in the way at present and bridge the gap between ideal protocols and practical implementations. In the framework, we are looking at the forward channel security rather than that of the backward one as the information reading totally depends on the states from the forward channel. If those states are kept secure, the security of the backward channel will be unquestioned naturally. In other words, if we reliably estimate the secrecy capacity of the forward channel, we are able to guarantee communication security by choosing the encoding strategy according to the secrecy capacity. Besides, inspired by the numerical security proof methods in QKD~\cite{Coles2015,Coles2017}, we resort to a numerical means of handling the analysis of the adversary's behavior instead of doing it manually. This could dramatically simplify the analysis process especially when we take into account the practical conditions, such as detector efficiency mismatch, side-channel effect, source imperfection and so on, in practical communications while some of the imperfections have been considered in QKD already such as those in references \cite{Liang2014,cao2015discrete}. It should be emphasized that this framework can be generalized to finite-size effect scenarios by using statistical methods and loosening the constraints used in our case. We are confident that this work would greatly propel the development in the QSDC field.

The rest of the paper is arranged as follows. In Sec.  \ref{sec-prototype}, we formally define the prototype of QSDC protocols and describe the communication process in quantum-mechanical language. Then, on the top of the prototype, the security analysis framework is constructed in Sec. \ref{sec-security}. Two optimization methods are proposed in Sec. \ref{sec-proposals} to meet various real-world needs and also the algorithm cores are both lined up in this part. Afterward, we apply our framework to several examples in Sec. \ref{sec-examples}. 
Then come the Conclusion and Appendix.

\section{General QSDC protocol}\label{sec-prototype}
\subsection{The protocol}
For simplicity of presentation, we will describe the entanglement based protocol while the prepare-and-measurement protocol can be viewed as an equivalent by the source replacement scheme~\cite{SR1992}.  

\textit{Step (1)}~The entanglement source (hypothetically held by Bob) allocates two qubits respectively to Alice and Bob. Repeat this for $N \ (N\rightarrow\infty)$ times.

\textit{Step (2)}~When Alice and Bob receive the qubits, Bob measures the qubit with his positive-operator valued measurements (POVMS) $\{F_j^B\}$ while Alice with probability $c\ll1$, measures by the POVMs $\{F^A_i\}$. At the meantime, they exchange the measurement outcome information via a classical channel and negotiate with each other to do a security estimation to make sure the quantum channel security capacity $\mathcal{C}_s$ is no less than 0. Otherwise, they abolish the communication and go back to step (1).

\textit{step (3)}~Alice encodes the rest $(1-c)N$ qubits with a certain set of unitary operators $\{U^A_k\}$ and resends those photons encoded to Bob and Bob decodes the message by using the measurement basis that he used in step (2) (if step (4) is needed, some check qubits are marked among the message qubits). So far a batch of secure communication has been completed. They go on to step (1) for the next round, or for the sake of robustness, they could additionally carry out step (4) even though no useful information would be leaked to the adversary.

\textit{step (4)}~Before decoding the message, Bob will do a second round check by measuring these in-advance inserted checking qubits from step (3) to guarantee the integrity of the information.

\subsection{Quantum-mechanical description of the prototype}
\begin{figure}
    \centering
    \includegraphics[width=1\linewidth]{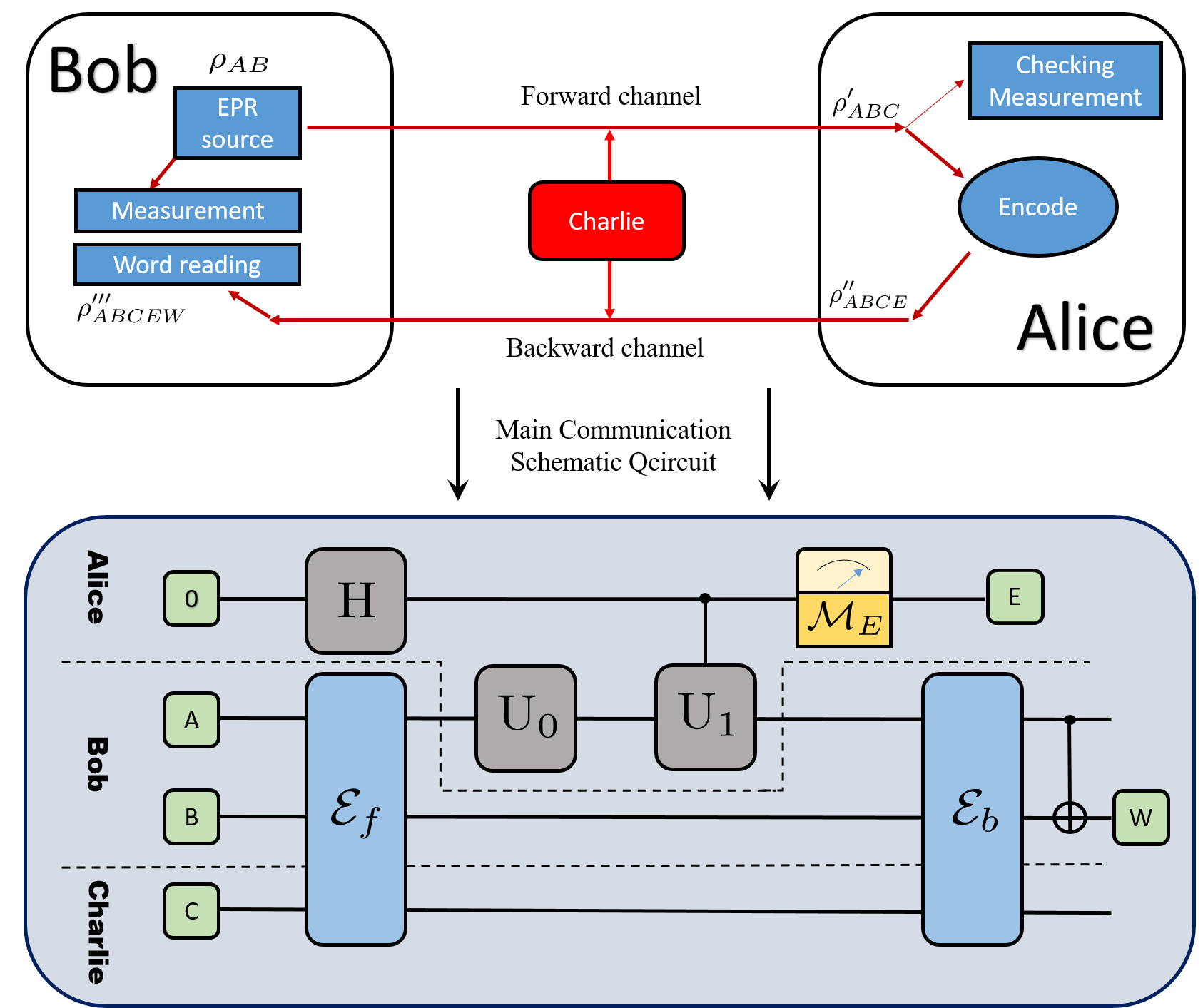}
    \caption{Schematic of quantum secure direct communication and the main communication quantum circuit. The lower part is used as an illustration of the main communication process. H is a Hardmard gate; ${\rm U_0}$ and ${\rm U_1}$ are the encoding unitary gates; $M_E$ is the post-selection measurement selected by Alice to encode classical information; $0$ denotes state $\ket{0}$ while A, B, C, E, W denote the corresponding registers: A, the qubit that Bob transmits to Alice; B, the qubit Bob possesses at his laboratory; C, the adversary's system (needless to be a qubit system); E, the register storing encoding information of Alice; W, the register storing Bob's decoding information. Here the entanglement state $\ket{\Phi^+}_{AB}=(\ket{00}+\ket{11})/\sqrt{2}$.}
    \label{fig-QSDC}
\end{figure}

The entanglement source produces a two-qubit state $\rho_{AB}$. Once the bipartite state (to be exact, the system of Alice) is exposed to the forward public quantum channel $\mathcal{E}_f$, it evolves into
\begin{equation}
\rho'_{ABC}=\mathcal{E}_f(\rho_{AB}),
\end{equation}which should be a pure state where the adversary Charlie holds the purifying system C since we suppose Charlie is powerful enough within the scope of quantum mechanics (while the recent discovery in reference~\cite{wen2021stable} bring an interesting phenomenon to light with respect to purification). After the encoding step, the whole system becomes 
\begin{equation}
    \rho''_{ABCE}=\mathcal{E}_E(\rho'_{ABC})
\end{equation}
with $\mathcal{E}_E(\cdot)$ an encoding map used to encode the message into the state and E as a register storing the encoding information. Here we are not going to specify the form of $\mathcal{E}_E(\cdot)$ as we will give the security proof without knowing the specific formula of $\mathcal{E}_E(\cdot)$.
As long as Alice has the states encoded, she resends them back to Bob who is going to do a word-reading map denoted by $\mathcal{E}_W(\cdot)$ where $W$ is the register system keeping the reading-out information. Thus comes the final compound state 
\begin{equation}
    \rho'''_{ABCEW}=\mathcal{E}_W\left(\mathcal{E}_b(\rho''_{ABCE}\mathcal)\right)
\end{equation}
with $\mathcal{E}_b(\cdot)$ as the backward channel. Similarly, the specific form of $\mathcal{E}_W$ is not important in the later analysis. The whole process description is illustrated as in Fig. \ref{fig-QSDC}. 

\section{Security proof framework}\label{sec-security}
According to information theory\cite{wyner1975wire}, secret communication can be guaranteed if the main channel capacity $\mathcal{C}_m$ of the legitimate bipartite users is bigger than that of the eavesdropping channel, $\mathcal{C}_c$, that's to say, the users can obtain a positive secrecy capacity
\begin{equation}
\begin{split}
    \mathcal{C}_s&= \mathcal{C}_m- \mathcal{C}_c\\
    &=I(E^A:W^B)-I(E^A:C)\\ 
    &=H(E^A|C)-H(E^A|W^B)\label{screcyCap2}
\end{split}
\end{equation}
where $I(X:Y)=S(\rho_{X})+S(\rho_Y)-S(\rho_{XY})$ represents mutual entropy and $H(X|Y)=S(\rho_{XY})-S(\rho_Y)$ represents the conditional entropy with $S(\rho)$ as the von Neumann entropy. The superscripts in the equations denote the possessors of the registers.

Similar to QKD's key rate analysis, to make sure the security of a QSDC protocol we have to consider the worst-case scenario when calculating the secrecy capacity, which means we think of
\begin{equation}
\mathcal{C}_s=\min[H(E^A|C)-H(E^A|W^B)]_{\rho'''_{ABCEW}}.\label{Cap-min}
\end{equation}
Note that the second term of the right hand side of Eq.~\eqref{Cap-min} is determined by Alice and Bob's error correction sacrifice. So to be more tight, it can be drawn out of the minimization, leaving
\begin{align}
\mathcal{C}_s&=\min[H(E^A|C)]_{\rho'''_{ABCE}}-H(E^A|W^B)_{\rho'''_{ABCEW}}\label{Cap-min-tight}\\
&\geq\min[H(K^B|C)-H(K^B|K^A)]_{\rho'_{ABC}}\nonumber\\
&~~~-H(E^A|W^B)_{\rho'''_{ABCEW}}\label{eq-ineq}\\ 
&=\min[H(K^B|C)]_{\rho'_{ABC}}-\gamma{h}(Q_f)-\gamma{h}(Q_b)\label{eq-concise}
\end{align}
where $K$ denotes an imaginary qubit-bit transforming map result for example in polarization system, $\ket{H},\ket{D}\rightarrow 0$ and $\ket{V},\ket{A}\rightarrow 1$ with $\ket{H},\ket{D},\ket{V},\ket{A}$ respectively stand for horizontal, diagonal, vertical, anti-diagonal polarizations. $\gamma$ is error correction rate. Without a further declaration, we will take $\gamma$ to be 1 as the error correction process is conducted at Shannon limitation for the following numerics.   Eq.~\eqref{eq-ineq} is derived from the fact that Charlie wouldn't know more useful information from the state $\rho''_{ABCE}$ than that from the forward channel eavesdropping since the encoding information depends totally on the original state of the qubits sent by Bob. The equal sign of Eq.~\eqref{eq-ineq} holds when Charlie reads out all the information from the qubits which he has controlled after forward channel taping. For the purpose of convenience, we define two terms to characterize the secrecy capacity (see Appendix \ref{A} for classified elaboration ). \textit{Secure capacity} $\mathcal{C}_s^s=\min[H(K^B|C)]_{\rho'_{ABC}}-H(K^B|K^A)_{\rho'_{ABC}}$. Under this capacity, the adversary knows nothing about the information sent. \textit{Reliable capacity} $\mathcal{C}_s^r$ stands for the secrecy capacity where backward channel error rate $Q_b$ and forward channel error rate $Q_f$ are both considered. For convenience, we take $Q_{f}=Q_{b}=Q$ to compute the reliable capacity since without extra influence caused by the adversaries, $Q_b$ would be no bigger than $Q_f$. In fact, considering the two-round compensation effect for the optical system~\cite{Plug&Play1998}, $Q_b$ should be always less than $Q_f$. Therefore, since $Q_{f}$ and $Q_{b}$ are both from observations, the ultimate goal of calculating the secrecy capacity is to optimize the first term of Eq.~\eqref{eq-concise},
\begin{equation}\label{eq-K}
  g=\min H(K^B|C)  
\end{equation}
with the other terms obtained from specific communication implementation. The qubit-bit map can also be visioned as an isometry $\mathcal{V}_K=\sum_l\kappa_l^B\otimes \ket{l}$ with respect to $\rho'_{AB}$, $\kappa_l^B$ being a projector subjected to $\sum_l\kappa_l^B=I_B$. Using that $\rho'_{ABC}$ is pure, we technically remove the dependence of Charlie's system in the optimization by the method mentioned in Refs.~\cite{Coles2011,Coles2012,Coles2015}, achieving
\begin{align}
g(\rho'_{AB})&=\min_{\rho'_{AB}}S(\rho'_{AB}||\sum_l\kappa^B_{l}\rho'_{AB}\kappa^B_l)\label{eq-optimization}\\
&{\rm s.t.}\  {\rm tr}(\rho'_{AB}\cdot F^A_i\otimes F^B_j)=Pr_{ij}\label{eq-constraint1}\\
&\hspace{2cm}{\rm tr}(\rho'_{AB})=1\\
&\hspace{2.5cm} \rho'_{AB}\succ0\label{eq-constraint3}
\end{align}
with $Pr_{ij}$ as the joint probability from observation of \textit{step (2)} of the protocol, where $S(\varrho||\varsigma)={\rm tr}(\varrho{\rm log}\varrho-\varsigma{\rm log}\varsigma)$ represents the relative entropy whose convexity over variable $\rho'_{AB}$ is guaranteed as is shown in~\cite{Watanabe2008}. In other words, $\mathcal{C}_s$ must have a global minimum over the feasible domain of a constrained density operator. Now the secrecy capacity is only relying on the composite system $\rho'_{AB}$ which can be easily constrained by the forward channel checking measurement. Notice that sometimes an imaginary post-selection is needed in general, that is,  this $\rho'_{AB}$ will be subjected to a post-selection map $\mathcal{G}$. This map won't impact the form of Eq.~\eqref{eq-optimization}, and more detailed discussion on this map could be found in Ref.~\cite{Coles2015}.

\section{Optimization proposals}\label{sec-proposals}
In this section, we are going to present two useful optimization methods to handle Eq.~\eqref{eq-optimization} in order to obtain the secrecy capacity. Beforehand, we define a feasible domain set $\mathcal{D}=\{\rho\succ0: {\rm tr}(\rho F^A_i\otimes F^B_j)=Pr_{ij},\ {\rm tr}(\rho)=1\}$ constrained by Eqs.~\eqref{eq-constraint1}-\eqref{eq-constraint3}.
Then, the optimization methods go as what follows.

\subsection{Special projected gradient descent}
First, we present a special projected gradient descent method (SPGD)~\cite{PGDqst,ICML} , in which, a "momentum" $\chi_s$ at $s-th$ iteration is involved to memorize the last sub-optimizing point. This method helps to avoid a dramatic descend and departing too much from the feasible domain $\mathcal{D}$ compared with the traditional gradient descent method. With $\mathcal{P}_{\mathcal{D}}(\cdot)$ as the map projecting any point in the density operator space into the feasible domain $\mathcal{D}$, the iteration core of the algorithm can be described as 
\begin{align}
    \chi_{s+1}&=\mu\chi_s-\zeta\cdot\nabla g(\rho_s),\\
    \rho_{s+1}&=\mathcal{P}_{\mathcal{D}}(\rho_s+\chi_{s+1}).
\end{align}
where $\mu$ controls the depth of the memorization of the last point and $\zeta$ is the step size which can be decided according to the practical iteration numbers or set to be a constant. $\nabla g(\rho_s)$ is the gradient of $g(\rho)$ in Eq.~\eqref{eq-optimization} when $\rho=\rho_s$ and $\rho_s$ is the $s-th$ iteration (sub-optimization) point. Empirically, this method works more properly than merely-projected gradient descent in our case considering the restriction to the feasible domain is kind of strong.
\subsection{Conditional gradient descent}
Also, we can apply the conditional gradient descent method (CGD)~\cite{FW2013} to the optimization in Eq.~\eqref{eq-optimization} as this method is talented for dealing with the optimization with constraints set in advance.  The main idea of the method is to transform an optimization problem into a series of linear optimizations until it finds a proper optimum. Based on this thought, the method works efficiently at the beginning interactions but converges slowly afterwords. The core part of the algorithm reads
\begin{align}
  \rho_{s+1}&=\zeta\omega_s+(1-\zeta)\rho_s,\\
  \omega_{s+1}&={\rm arg}\max_{\sigma\in\mathcal{D}}{\rm tr}(\nabla g(\rho_s)\cdot\sigma),
\end{align} 
where $\zeta$ also denotes the step size which can be decided by another minimization in each iteration to make sure an optimal step decrease, or simply determined by the iteration number as the former method does. As a rough approximation has been made in each sub-optimization, finding the ultimate optimum will come across a precision problem. Usually, the global optimum stands outside the feasible domain leaving the constrained optimum lying on the boundary of the constraints. This might also pose a numerical challenge for the "approximation" optimization because the behaviour of it is kind of subtle around the boundary.  

\section{Applications to specific examples}\label{sec-examples}
With all the framework defined and optimization methods proposed, we then apply our security analysis approach to several protocols where some are hard (or even impossible) to achieve an analytical security proof, such as those with all detector efficiencies included. 

\subsection{DL04 protocol and DL04-6-state protocol}
\begin{figure}
    \centering
    \includegraphics[width=1\linewidth]{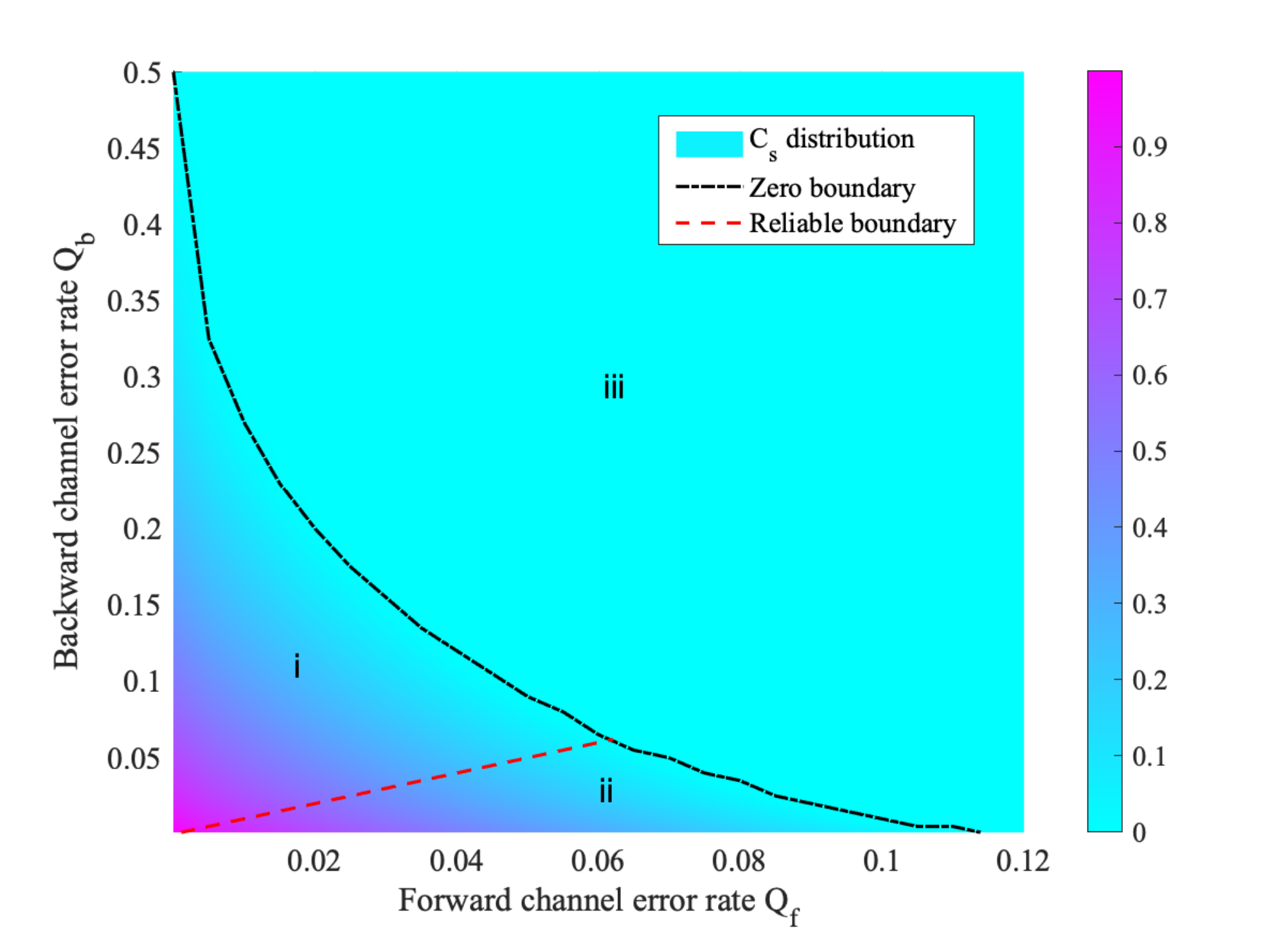}
    \caption{Secrecy capacity distribution of DL04 protocol vs forward channel error rate $Q_f$ and backward channel error rate $Q_b.$ The black dash line is the boundary of the secure and insecure scenarios. "iii" denotes the insecure one while "i"+"ii" represents the opposite. The red dash line represents the boundary where $Q_f=Q_b$ that partitions the part of secure scenario.}
    \label{fig-DL04color}
\end{figure}

\begin{figure}
    \centering
    \includegraphics[width=1\linewidth]{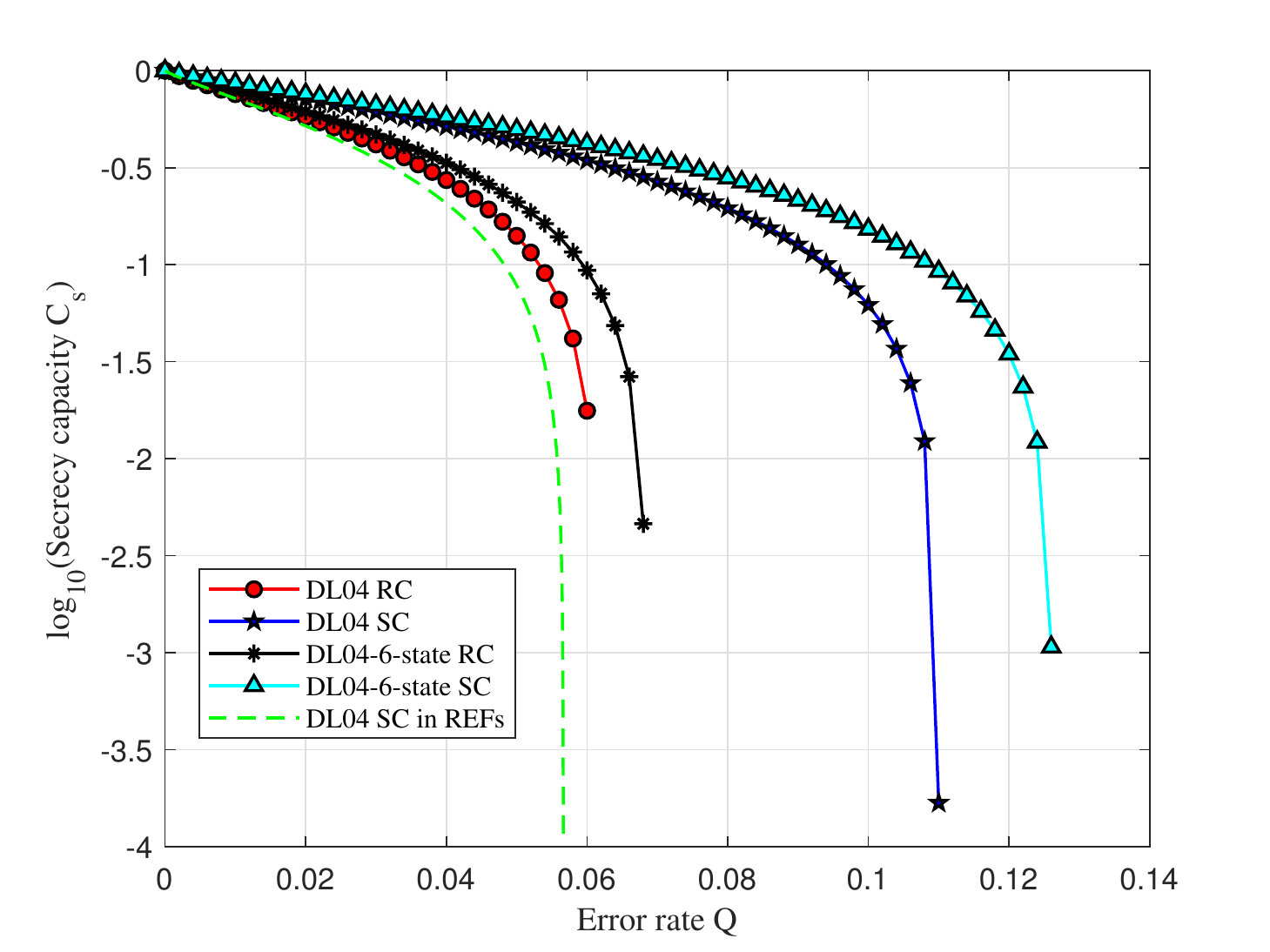}
    \caption{Secrecy capacity subjected to logarithm based on 10 vs error rate Q. All the capacities stand for DL04 protocols classical or improved. The green dash line denotes the result from Refs.~\cite{wu2019security,qi2019implementation} while the others are derived from the new numerical framework. The abbreviation "RC" represents reliable capacity while "SC" represents secure capacity. Every symbol here denotes a numerical result. Note that when $Q_f$ and $Q_b$ are used together, we take them both as Q, i.e., $Q_f~=~Q_b$ to facilitate the plotting and demonstration.}
    \label{fig-CsCompare}
\end{figure}
First, as an appetite try-on, we utilize the new framework to calculate the secrecy capacity of the famous QSDC protocol DL04~\cite{deng2004secure} based on entanglement source. According to the source replacement scheme, both entanglement-based and prepare-and-measure protocols can be equalized. The result of the secrecy capacity vs forward and backward channel error rates, $Q_f$ and $Q_b$ is shown in  Fig. \ref{fig-DL04color} where three partitions denoted by i, ii, and iii are divided by two boundaries, respectively zero capacity boundary and reliable capacity boundary. The black curve seems a bit defective because of numerical precision. This can be refined by tightening the precision parameters and increasing the dot density. In Fig. \ref{fig-CsCompare}, we compare the secrecy capacities derived from the new method and the previous method in Refs.~\cite{wu2019security,qi2019implementation}. 
Our new method beats the previous one for both secure capacity and reliable capacity. We also make some variation on the classical DL04 protocol via introducing $\sigma_y$ basis checking measurement when carrying out the security checking phase while more general checking mode could be considered like having been shown in Ref.~\cite{ZY1}. That is, in the modified protocol, DL04-6-state protocol, more information can be obtained from the check phase used to bound the adversary's knowledge of the state shared by Alice and Bob. As demonstrated in the figure, this modification improves capacity for it shrinks the searching space of the problem Eq.~\eqref{eq-optimization}.

\begin{figure}
    \centering
    \includegraphics[width=1\linewidth]{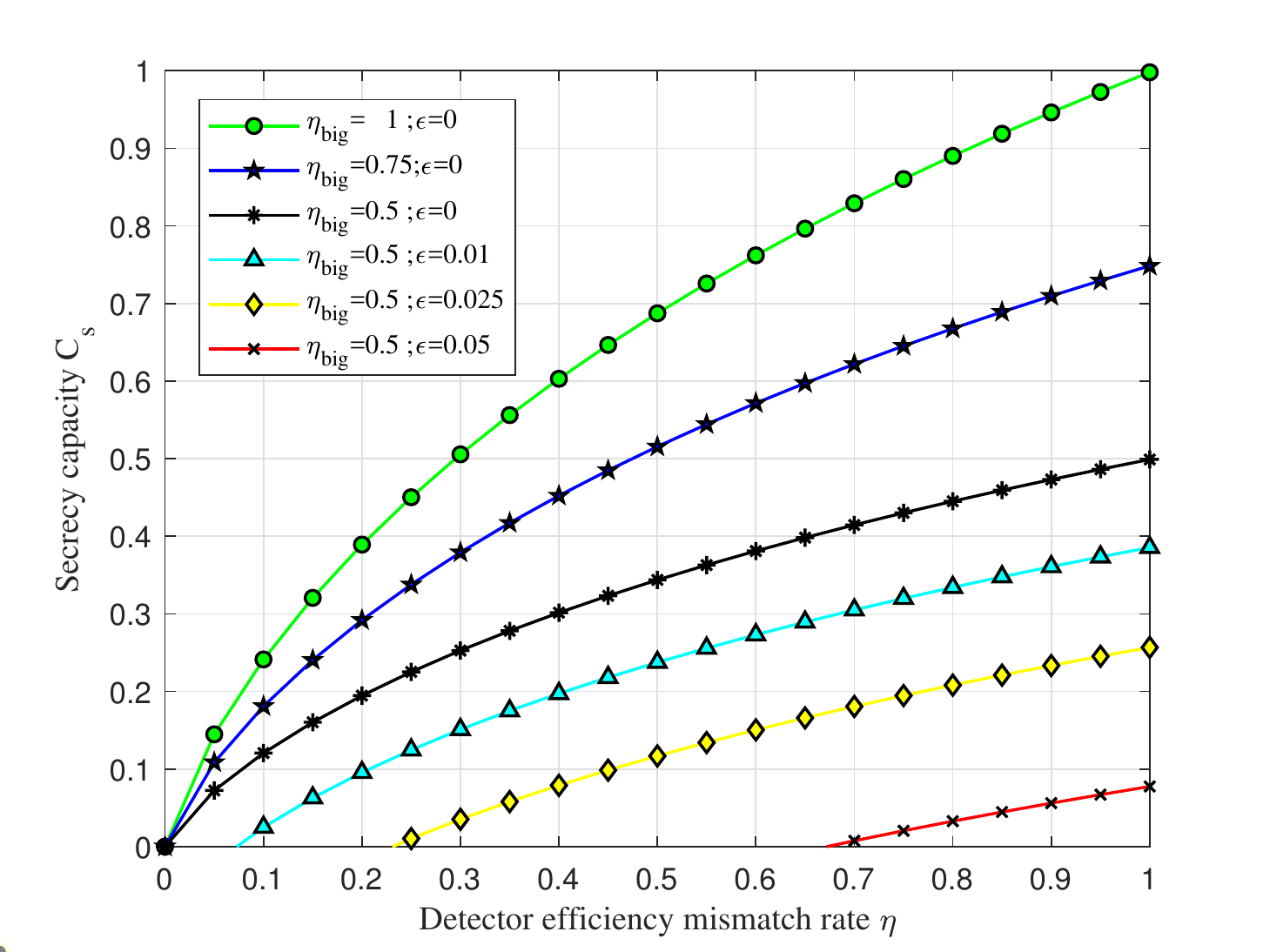}
    \caption{Secrecy capacity $C_s$ vs detector efficiency mismatch rate $\eta$. The bigger detector efficiency denoted by $\eta_{big}$ while the smaller one is $\eta\cdot\eta_{big}$. The depolarizing channel parameter $\epsilon$ varies in $(0, 0.01, 0.025, 0.05)$ and $\eta_{big}$ varies in $(1, 0.75, 0.5)$. Note that the secrecy capacities here are referred to as reliable capacities. Every symbol denotes a numerical result.}\label{fig-mismatch}
\end{figure}

\subsection{Imperfection of detectors}\label{Imperfection}
In practical communication, the optical detectors are far from perfect as the real-world efficiencies of the detectors are not 1. Meanwhile, each of the detectors used in the experiment may not match one another, i.e., they possess different efficiencies. If every detector matches, one can simply attribute the common loss rate of the detector to the channel loss, which would decrease the capacity proportionally. However, the mismatch of the detectors can not be handled by this trivial attribution since the adversary may take advantage of the loophole caused by the spatial-mode detector-efficiency mismatch~\cite{mismatch2010,mismatch2015}. So it poses a problem to be considered in the implementation of QSDC. Under our framework, this problem can be easily addressed by incorporating each of the efficiencies into the checking measurement operators. Note that the mismatch of Bob's decoding detectors does not ruin the security.

Considering above, we apply our framework to the analysis of detector efficiency mismatch cases. In order to obtain a set of experimental data, we simulate the measurement results under depolarizing channel $\mathcal{E}^d$, that is,
\begin{equation}\label{eq-depo}
\mathcal{E}^d(\rho_{AB})= 
\frac{\epsilon}{d_B(d_A-1)}\begin{pmatrix}
I_A^{\notin0}&\\
& 0
\end{pmatrix}\otimes I_B+(1-\epsilon)\rho_{AB}
\end{equation}
where $\epsilon$ is the depolarizing parameter. $d_A$ and $d_B$ are the dimensions of respectively Alice's and Bob's systems. In the simulation, we vision Bob's detectors as ideal ones as it should be in the prepare and measurement scenario while Alice's are imperfect. "$\notin0$" denotes the space except the non-detection subspace (or called vacuum space). It should be emphasized that this framework can be used under arbitrary quantum channels including but not limited to the depolarizing one. For comparison, we set the bigger detector efficiency varying in $(1, 0.75,0.5)$ and tune the mismatch rate $\eta$ semi-continuously to observe the reliable capacity at each circumstance. From Fig. \ref{fig-mismatch}, the detector efficiency mismatch will certainly ruin the secrecy capacity of QSDC. Especially, we calculate a family of lines of $\eta_{big}=0.5$ for these detector settings are close to practical ADP detectors, so the result may be used as a reference to real cases.

\subsection{Comparison of the optimization methods}
\begin{figure}
    \centering
    \includegraphics[width=1\linewidth]{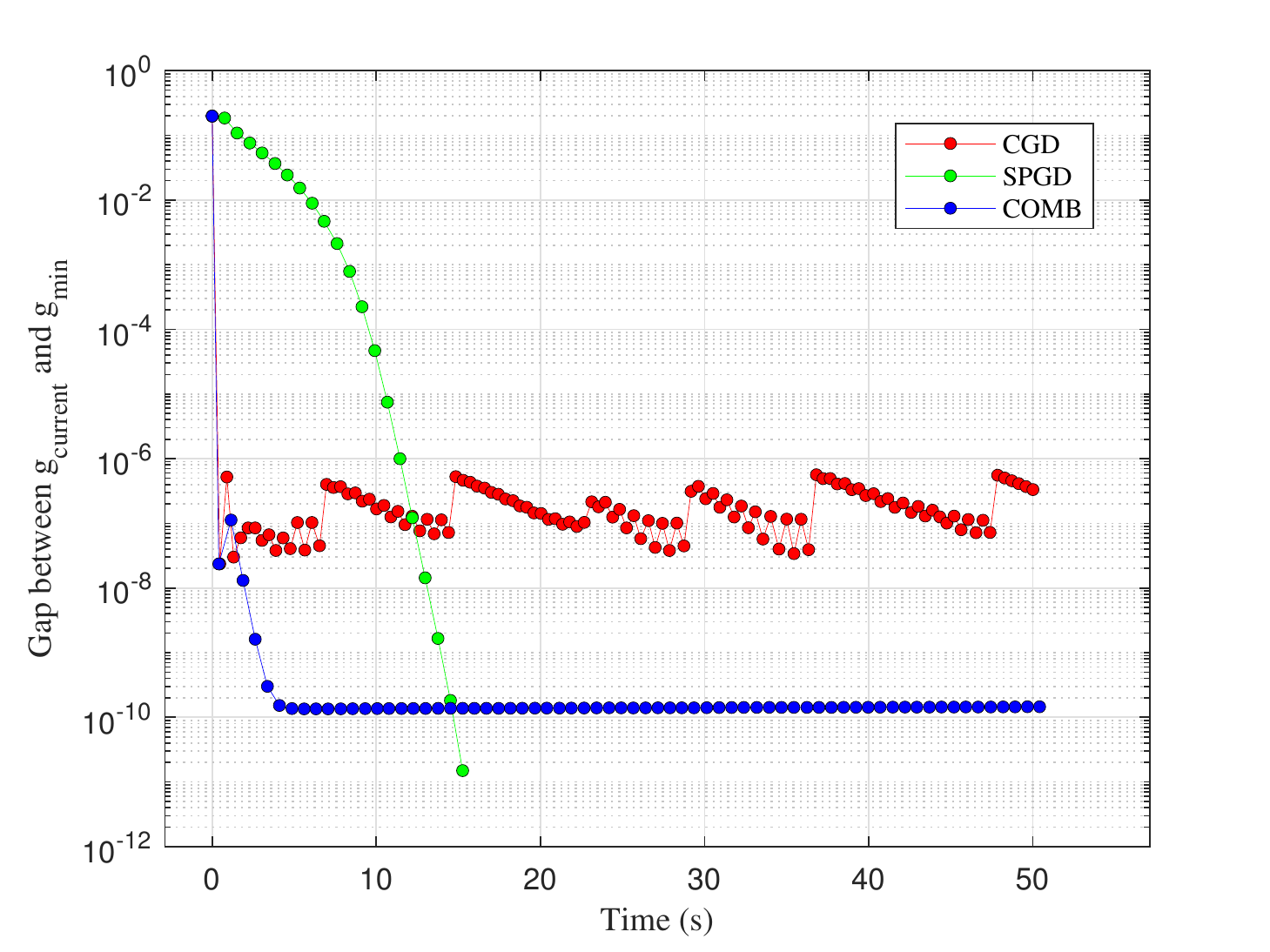}
    \caption{The gap between current objective function value $g_{current}$ and the final minimum $g_{min}$. The dots on each line denote the iteration points. The green dotted line gives the optimization trend of special projected gradient descent (SPGD) method while the red line shows that of the conditional gradient descent (CGD) method. The blue line demonstrates the trend of the method stemming from the combination of CGD and SPGD. The comparison data are acquired under DL04 protocol background.}\label{fig-time}
\end{figure}
As in Sec. \ref{sec-proposals}, we have brought forward two optimization methods, SPGD and CGD. In this part, we compare the speeds and optimizing depths of the two methods under DL04 protocol framework to illustrate their properties when solving the problem Eq.~\eqref{eq-optimization}. In Fig. \ref{fig-time}, the relations between optimizing depth and the time used to reach this depth are plotted. The optimizing depth is characterized by the gap between current sub-optimization value of the function $g(\rho)$ and the finial optimum which is fixed in advance according to SPGD's limit depth. Judging from the figure, we find that the SPGD method goes deeper and deeper in every iteration and eventually reaches the final "deepest" minimum illustrated as the green dots. The red dotted line shows CGD reaches a favourable sub-minimum 
in a very short time but it is hard for CGD to achieve a high precision result and after the first very efficient iteration, it oscillates back and forth around the first depth. Then it goes even worse after a few iterations. In conclusion, both of the two methods possess their advantages. To take advantage of each method, we combine them together as a complementary one (COMB) whose performance is demonstrated as the blue dotted line. This combination cuts down the ruining time to achieve an appropriate minimum up to the precision of $~10^{-10}$ and considerably save half of the time of SPGD. Note that in the literature~\cite{Coles2017}, the authors propose a dual problem of the optimization to make sure the tightness of the results derived from numerics. That is a good choice to guarantee the numerical results but it truly perplexes the problem itself. And sometimes when the requirement of the precision is pretty high, this dual optimization fails as shown in Fig. 2 and Fig. 7 in Ref.~\cite{Jie-PRX}. We propose these three methods as choices to make sure the optimization goes deep enough so that we could reliably keep the first significant digits of the numerical results.
\section{Conclusion}
We have established a new security analysis framework oriented for quantum secure direct communication. First of all, the prototype of a generic QSDC protocol is redefined, and following this prototype we present the framework quantum-mechanically. Furthermore, we investigate the security of different variations of DL04 protocol via the new framework driven by numerical optimizations. Meanwhile, pursuing preciser and faster optimization, we have proposed two methods SPGD and CGD and studied their properties. As a result of the comparison, one could choose these methods according to practical requirements. Above all, we remark that this framework can be used to analyse almost any practical QSDC protocols as it simplifies the 
investigation of the adversary's actions and can take into account the implementation conditions such as real-world detector efficiencies and the imperfection of the communication source. With the constructive advantages of the framework, it can be extended to the finite-size secrecy capacity analysis as well. All in all, this framework may open up a broad avenue for the development of QSDC among the research community. 

\acknowledgements{We would like to thank Jiawei Wu for his generous providing of the comparison data in Fig. \ref{fig-CsCompare} and thank Jie Lin for the help of the numerical techniques. This work was supported by National Key Research and Development Program of China under Grant No.2017YFA0303700, Key Research and Development Program of Guangdong province under Grant No.2018B030325002, National Natural Science Foundation of China under Grants No.11974205, and Beijing Advanced Innovation Center for Future Chip (ICFC).}

\appendix
\section{Definitions and abbreviations} \label{A}
Secrecy capacity labeled by $\mathcal{C}_s$: the difference of the main channel capacity and the tap channel capacity.

Secret capacity (SC) labeled by $\mathcal{C}^s_s$: The secrecy capacity when backward channel is not considered. As described in the main text, the secrecy of QSDC can be totally guaranteed by forward channel checking, i.e., if $\mathcal{C}_s^s ~>~0$, the communication is secure.

Reliable capacity (RC) labeled by $\mathcal{C}_c^r$: The secrecy capacity when both forward and backward channels are considered. In addition to guaranteeing the secrecy of QSDC, if $\mathcal{C}_s^r~>~0$, the integrity of the information conveyed during the communication is guaranteed.

\section{The derivation of the main optimization problem}
In this section, we are going to derive the main optimization problem in Eq.~\eqref{eq-optimization} from Eq.~\eqref{eq-K}.
\begin{equation}
    g=\min H(K^B|C)=\min[ S(\rho'^*_{CK^B})-S(\rho'_C)]
\end{equation}
Using that $\rho'_{ABC}$ is pure and $\mathcal{V}_K=\sum_l\kappa_l^B\otimes\ket{l}$ is an isometry, we obtain
\begin{align}
    g&=\min \{S[{\rm tr}_{CK^B}(\rho'^*_{ABCK^B})]-S(\rho'_{AB})\}\\
    &=\min \{S[{\rm tr}_{CK^B}(\sum_l\kappa_l^B\otimes\ket{l}\rho'_{ABC}\sum_{l'}\kappa_{l'}^B\otimes\bra{l'})]\nonumber\\
    &~~~~~~~~~-S(\rho'_{AB})\}\\ 
    &=\min\{S[{\rm tr}_C(\sum_l\kappa_l^B\rho'_{ABC}\kappa_l^B)]-S(\rho'_{AB})\}\\
    &=\min\{S(\sum_l\kappa_l^B\rho'_{AB}\kappa_l^B)-S(\rho'_{AB})\}\\
    &=\min\{-\sum_l{\rm tr}[\kappa_l^B\rho'_{AB}\kappa_l^B{\rm log}(\sum_{l'}\kappa_{l'}^B\rho'_{AB}\kappa_{l'}^B)]\nonumber\\
    &~~~~~~~~~-S(\rho'_{AB})\}\\
    &=\min\{-{\rm tr}[\rho'_{AB}{\rm log}(\sum_l\kappa_{l}^B\rho'_{AB}\kappa_{l}^B)]-S(\rho'_{AB})\}\\
    &=\min S(\rho'_{AB}||\sum_{l}\kappa_{l}^B\rho'_{AB}\kappa_{l}^B)
\end{align}
\section{Entanglement based DL04 protocol with detector efficiency mismatch}
We establish the model for entanglement based DL04 protocol with detector efficiency mismatch in this part. The POVMs Alice's measurement can be expressed as
\begin{align}
&F^A_1=p_z\eta_{big}\ket{0}\bra{0}\oplus(0)^{\in0}\\
&F^A_2=p_z\eta_{big}\eta\ket{0}\bra{0}\oplus(0)^{\in0}\\
&F^A_3=(1-p_z)\eta_{big}\eta\ket{+}\bra{+}\oplus(0)^{\in0}\\
&F^A_4=(1-p_z)\eta_{big}\eta\ket{-}\bra{-}\oplus(0)^{\in0}\\
&F^A_5=I-\sum_{j=1}^4F_j^A.
\end{align}
where $\ket{0},\ket{1}$ are the basis vectors of the Pauli operator $\sigma_z$, $\ket{+}, \ket{-}$ are the basis vectors of $\sigma_x$ and $(0)^{\in0}$ is a 1-by-1 "matrix" in non-click subspace. Similarly, the POVMs for Bob's measurement are 
\begin{align}
&F^B_1=p_z\ket{0}\bra{0},\\ &F^B_2=p_z\ket{1}\bra{1},\\
&F^B_3=(1-p_z)\ket{+}\bra{+},\\
&F^B_4=(1-p_z)\ket{-}\bra{-},
\end{align}
as his detectors are viewed as ideal ones in order to completely model the original DL04 protocol which utilizes single photons in the scheme. $p_z$ denotes the $\sigma_z$-basis-choosing factor. For simplicity of processing, $p_z$ should be very close to 1 or 0 alternatively. Otherwise, a normalization factor has to be introduced in order not to underestimate the secrecy capacity as after the forward channel in the protocol, we assume an imaginary qubit-bit map to evaluate the information amount. As a mater of fact, there is no basis choosing phase during the formal communication period except the checking phase. Specifically in our nurmerics, we set $p_z=0.999$. The simulated date used in Sec. \ref{Imperfection} are produced as
\begin{equation}
    Pr_{ij}={\rm tr}(\mathcal{E}^d(\rho_{AB}) F^A_i\otimes F_j^B).
\end{equation}
$\mathcal{E}^d(\cdot)$ is defined as in Eq.~\eqref{eq-depo}.
The post-selection map $\mathcal{G}$ can be described by two Kraus operators $\{\mathcal{K}_1, \mathcal{K}_2\}$. We further choose
\begin{align}
    \mathcal{K}_1=(\ket{0}_K\otimes\sqrt{F^A_1}+\ket{1}_K\otimes\sqrt{F^A_2})\otimes\sqrt{F^B_1+F^B_2},\\
    \mathcal{K}_2=(\ket{0}_K\otimes\sqrt{F^A_3}+\ket{1}_K\otimes\sqrt{F^A_4})\otimes\sqrt{F^B_3+F^B_4}    
\end{align}
so that the projector operators in Eq.~\eqref{eq-optimization} reads \begin{align}
 &\kappa_0=\ket{0}_K\bra{0}\otimes I_{AB}\\
 &\kappa_1=\ket{1}_K\bra{1}\otimes I_{AB}.
\end{align} 
Note that $\kappa_l$ here is no longer in terms of the original systems, A and B. With all the setting listed above, Fig. \ref{fig-mismatch} in Sec. \ref{Imperfection} should be achieved through the numerics.

%%%%%%%%%%%%%%%%%%%%%%%%%%%%%%%
%%%%%%%%%%%%%%%%%%%%%%%%%%%%%%%

\bibliographystyle{plain}

\end{document}